\documentclass[11pt]{article}

\usepackage{amssymb} 
\usepackage{amsmath}     
\usepackage{amsthm}
\usepackage{todonotes}
\usepackage{mathtools}
\usepackage{braket}

\usepackage[ruled,vlined,linesnumbered]{algorithm2e}
\SetKw{KwNot}{not}
\SetKw{KwOR}{or}
\SetKw{KwAND}{and}
\SetKw{KwExitFor}{exit for-loop}

\usepackage{authblk}

\usepackage{a4wide}
\usepackage{graphicx}
\usepackage{hyperref}
\newcommand{\Rset}{\mathbb{R}}
\newcommand{\Xset}{\mathbb{X}}

\newtheorem{prop}{Proposition}

\newtheorem{ass}{Assumption}

\begin{document}

\title{Soft robust solutions to possibilistic optimization problems}

\author[1]{Adam Kasperski}
\author[1]{Pawe{\l} Zieli\'nski}

\affil[1]{
Wroc{\l}aw  University of Science and Technology, Wroc{\l}aw, Poland\\
            \texttt{\{adam.kasperski,pawel.zielinski\}@pwr.edu.pl}}

 \date{}

\maketitle

\begin{abstract}
This paper discusses a class of uncertain optimization problems, in which unknown
parameters are modeled by fuzzy intervals. The membership functions of the fuzzy intervals are interpreted as possibility distributions for the values of the uncertain parameters. It is shown how the known concepts of robustness and light robustness, for the traditional  interval uncertainty representation of the parameters,  can 
 be generalized to choose solutions that  optimize against plausible  parameter realizations
  under the assumed model of uncertainty in the possibilistic setting.
  Furthermore, 
  these solutions can be computed efficiently for a wide class of problems, in particular for linear programming problems with fuzzy parameters in  constraints and objective function.
 Thus the problems under consideration are  not much computationally harder than their deterministic counterparts.
   In this paper a theoretical framework is presented and results of some computational tests are shown.
\end{abstract}


\section{Introduction}
In this paper we investigate the following optimization problem with uncertain parameters:
\begin{equation}
\widetilde{\mathcal{P}}: \begin{array}{lll}
 \min &\pmb{c}^T\pmb{x}\\
 \text{s.t.} &\widetilde{\pmb{A}}\pmb{x} \leq \pmb{b}\\
 		& \pmb{x}\in \Xset\subset \Rset_+^n
  \end{array}
 \label{lpf}
\end{equation}
In formulation~(\ref{lpf}), $\pmb{x}$ is an $n$-vector of decision variables, $\widetilde{\pmb{A}}=(\tilde{a}_{ij})$ is an $(m\times n)$-matrix of imprecise constraint coefficients and $\pmb{c}$ is an $n$-vector of objective function coefficients. The meaning of $\widetilde{\pmb{A}}$ and the relation $\leq$ depends on the model of uncertainty assumed
and it will follow from the context.
For simplicity of presentation, we first assume that the vector $\pmb{c}$ of the objective function coefficients is precisely known. We will show later, in Section~\ref{secfobj}, that the approach proposed in this paper can be easily extended to the case of uncertain objective function coefficients~$\tilde{\pmb{c}}$.
An $m$-vector $\pmb{b}$ of right hand sides is also assumed to be precisely known. It does not cause loss of generality, as we can always add artificial variables and include uncertain right hand sides in matrix $\widetilde{\pmb{A}}$ 
(see, e.g.,~\cite{BS03}).
 We will denote by $\tilde{\pmb{a}}_i^T\pmb{x}\leq b_i$  the $i$th imprecise constraint in~(\ref{lpf}), where $\tilde{\pmb{a}}_i^T$, $i\in[m]$, is the $i$th row of $\widetilde{\pmb{A}}$  (throughout the paper we will use the notation $[m]=\{1,\dots,m\}$). 
 Set~$\Xset$ is a bounded subset of $\Rset_+^n$, where $\Rset_+$ is the set of nonnegative reals. For example, if $\Xset$ is a bounded polyhedron, then we get an  \emph{uncertain linear programming problem.}
 If $\Xset\subseteq\{0,1\}^n$ ($\Xset$ is a finite set), then~(\ref{lpf}) becomes an uncertain combinatorial optimization problem. For  a particular realization of the constraint coefficients $\pmb{A}\in \Rset^{m\times n}$ (called \emph{scenario}), we get a 
 deterministic
 counterpart $\mathcal{P}$ of $\widetilde{\mathcal{P}}$, which is a traditional optimization problem.

A typical method of solving~(\ref{lpf}) consists in replacing the imprecise constraints with some crisp equivalents and solving the resulting mathematical programming problem (see,~e.g.,~\cite{BN09,EV16,KM05,LH92,PZT11,S93,ST90}). The method of constructing such a problem depends on the interpretation of the imprecise parameters, which in turn, depends on the information available. In many cases the resulting model is harder to solve than the deterministic counterpart of~(\ref{lpf}).

If $\tilde{\pmb{a}}_i$, $i\in [m]$, are vectors of random variables with known probability distributions, then 
\emph{stochastic optimization framework} can be used (see, e.g.,~\cite{KM05}). Namely,  we can replace the imprecise  constraints in~(\ref{lpf})
with chance constraints of the form 
$${\Pr}(\tilde{\pmb{a}}_i^T\pmb{x}\leq b_i)\geq 1-\epsilon_i,$$
where $\epsilon_i\in (0,1)$ is a given risk (significance) level. 
In practice, however, it is often difficult or even impossible to provide the parameter distributions. Furthermore, the resulting problem with chance constraints can be hard to solve~\cite{KM05}.

If the probabilistic information about the parameters is not available, then \emph{robust optimization framework} can be applied (see, e.g.,~\cite{BN09}, \cite{BN99}). Suppose we only know that $\pmb{A}=(a_{ij})\in \mathcal{U}\subseteq \Rset^{m\times n}$, where $\mathcal{U}$ is a given \emph{uncertainty (scenario) set}, containing all possible realizations (scenarios) of the uncertain constraint coefficients. Using the robust framework, problem~(\ref{lpf}) is then expressed as:
\begin{equation}
 \begin{array}{lll}
\min &\displaystyle \pmb{c}^T \pmb{x}\\
  \text{s.t.}	      & \pmb{A}\pmb{x}\leq \pmb{b} &  \forall \pmb{A} \in \mathcal{U}\\
 		& \pmb{x}\in \Xset
  \end{array}
 \label{lpr}
\end{equation}
Solutions to~(\ref{lpr}) (if they  exist), called \emph{strictly robust},  can be very conservative, as we require that the constraints are satisfied for all possible realizations of the parameters (see~\cite{S73}).  Several methods of relaxing the strict robustness have been proposed in the existing literature. One of the most common was introduced in~\cite{BS04}, where it is assumed for each constraint, that only a subset of the imprecise parameters can take their worst values. Then, each constraint is satisfied with a reasonable probability. We will describe this idea in more detail in Section~\ref{secrob}.

Another method of softening~(\ref{lpr}) is to relax the right hand sides of the constraints, which leads to the concept of \emph{light robustness}, originally proposed in~\cite{FM09} and further discussed in~\cite{S14}.  In typical situations, where everything goes smoothly without any disturbances, the constraint coefficients will take some \emph{nominal values} $\widehat{\pmb{A}}=(\hat{a}_{ij})\in \mathcal{U}$, where $\hat{a}_{ij}$ is the nominal value of
 uncertain coefficient~$\tilde{a}_{ij}$.
 A robust solution should be feasible in the nominal scenario and also not too far from optimality under this scenario. This can be modeled by adding the crisp constraints $\widehat{\pmb{A}}\pmb{x}\leq \pmb{b}$ and $\pmb{c}^T\pmb{x}\leq \hat{c}+\rho_0$, where $\hat{c}$ is the optimal objective value of~(\ref{lpf}) under the nominal scenario $\widehat{\pmb{A}}$ and $\rho_0\geq 0$ is a fixed tolerance. Finally,  the constraints should be satisfied for all scenarios with some possible tolerances (deviations). The goal is now to minimize a distance of the deviations to the zero-vector. The light-robustness counterpart of~(\ref{lpf}) takes then the following form~\cite{FM09, S14}:
\begin{equation}
 \begin{array}{lll}
\min &\|\pmb{\gamma}\|\\
  \text{s.t.}	      
  		&\pmb{A}\pmb{x}\leq \pmb{b}+\pmb{\gamma} &  \forall \pmb{A} \in \mathcal{U}\\
		& \widehat{\pmb{A}}\pmb{x}\leq \pmb{b}\\
		& \pmb{c}^T\pmb{x}\leq \hat{c}+\rho_0\\
		& \pmb{\gamma}\geq \pmb{0}\\
 		& \pmb{x}\in \Xset
  \end{array}
 \label{lplr}
\end{equation}
where $\|\cdot\|$ denotes a given norm and $ \pmb{\gamma}\in \Rset^m_{+}$ is a vector of $m$-decision variables,
slack variables, 
that  take  strictly positive values if the corresponding  constraints are violated.

 In the  classical  stochastic approach a full probabilistic information about the problem parameters is available, while in the traditional robust approach we may only know the supports of the distributions of the random parameters. Many problems arising in practice are located  between these two boundary cases. Namely, a partial information about parameter distributions, such as their mean (nominal) values and variances, is available. We can then seek solutions that hedge against the worst probability distributions which may appear. This leads to various robust distributionally models discussed, see for instance~\cite{DY10, GS10}. 
Another method of modeling incomplete probabilistic information involves fuzzy sets with their possibilistic interpretation. Namely, we can assume that $\tilde{\pmb{a}}_i$, $i\in [m]$, are vectors of fuzzy quantities with specified \emph{possibility distributions}. Possibility distribution can be seen as an estimation (upper bound) on the unknown probability distribution and some methods of constructing it from the available data can be found in~\cite{DD06, DP88}. We can now utilize this additional possibilistic information to improve the solution robustness, by using possibility and necessity measures. For example, we can replace 
the imprecise constraints of~(\ref{lpf}) with fuzzy chance constraints of the form 
\[
{\Pi}(\tilde{\pmb{a}}_i^T\pmb{x}\leq b_i)\geq 1-\epsilon_i \text{ or } {\rm N}(\tilde{\pmb{a}}_i^T\pmb{x}\leq b_i)\geq 1-\epsilon_i,
\]
 where $\Pi$ and ${\rm N}$ are possibility and necessity measures, respectively (see, e.g.,~\cite{IR00, LIU01, PZT11}). 
For a deeper discussion on various approaches  used in fuzzy optimization
we refer the reader to~\cite{IIK92, IRTV03, LK10,LH92,RV02,R06,ST90}.

 The aim of this paper is to extend the robust concepts proposed in~\cite{BS04, FM09, S14} to the fuzzy case in the possibilistic setting. As in~\cite{BS04}, we will assume that for each uncertain parameter (matrix coefficient)~$\tilde{a}_{ij}$ an interval of possible values is provided, which is symmetric around its nominal value~$\hat{a}_{ij}$. This value is usually chosen as the most likely one. Indeed,
in practice, 
knowledge about uncertainty  of a parameter is usually expressed as
a possible deviation ($\pm \overline{a}_{ij}$) from $\hat{a}_{ij}$, which 
means that the actual parameter
will take some value within the interval $[\hat{a}_{ij}-\overline{a}_{ij},\hat{a}_{ij}+\overline{a}_{ij}]$, but it is not possible at present to
predict which one. In consequence, it induces a 
 simple \emph{interval uncertainty representation}  (see, e.g.,~\cite{KY97}).
 In our approach a possibility distribution within this interval can also be prescribed. This possibility distribution can be seen as an upper bound on the unknown probability distribution (see, e.g.,~\cite{DD06,DFMP04}). Now, some parameter values within this interval are more plausible than others, which extends and refines the traditional interval uncertainty representation. 
 Following~\cite{BS04}, we make a reasonable assumption that in practical situations it is unlikely that all parameters will deviate from their nominal values at the same time. Accordingly, we specify at most how many coefficients in each constraint can deviate from their nominal values. Then, following~\cite{FM09, S14}, we provide an acceptable increase in the cost of a solution found. In order to choose a robust solution, we propose two necessity measure based criteria. Using the first criterion we seek a solution, called a \emph{best necessarily feasible}, for which we are sure with the highest degree that it is protected against the worst parameter realizations. The second criterion, called a \emph{best necessary soft feasibility}, is a relaxation of the previous one and is similar in spirit to the idea of light robustness (see model~(\ref{lplr})). It is worth pointing out that both criteria will lead to computationally tractable problems for some important special cases of~(\ref{lpf}). 
 
The following natural assumption will be needed throughout the paper. 
\begin{ass}
 Set $\Xset$ is a nonempty bounded subset of~$\Rset^n_{+}$ and there exists $\pmb{x}\in \Xset$, feasible to
 $\widehat{\pmb{A}}\pmb{x}\leq \pmb{b}$, where 
 $\widehat{\pmb{A}}=(\hat{a}_{ij})$ is a matrix of the nominal constraint coefficient values.
 \label{assf}
\end{ass}
 The above assumption ensures that all the programs (models) proposed in this paper are bounded and
  feasible, when the parameters are precise.

 This paper is organized as follows. In Section~\ref{secrob} we recall the concepts of robustness and light robustness proposed in~\cite{FM09, BS04, S14}. In Section~\ref{secpos} we apply possibility theory to model the uncertain problem parameters. We introduce a possibilistic model of uncertainty and provide its interpretation. In Section~\ref{secsfuzzrob} we propose a concept of choosing a solution, which extends the traditional robust approach to the fuzzy (possibilistic) case. In Section~\ref{secsoftrob} we further generalize the concept from Section~\ref{secsfuzzrob} by using the idea similar to light robustness.  In Section~\ref{secfobj} we  show how the uncertain objective function can be considered in our model. In Section~\ref{secsol} we provide an algorithm for solving the problem and identify special cases which can be solved in polynomial time.
 Finally, in Section~\ref{secexper} we show results of some experiments, which suggest that taking additional information about the uncertain parameters into account may lead to  solutions
 with a better quality 
 over a set of plausible  parameter realizations.
 
\section{Robust and light robust solutions under interval uncertainty}
\label{secrob}

In this section we briefly recall the robust and light robust approaches proposed in~\cite{BS04, S14, FM09}.  Consider the $i$th imprecise constraint $\tilde{\pmb{a}}_i^T\pmb{x}\leq b_i$.  Suppose that $\tilde{a}_{ij}$, $j\in [n]$, is a random variable, symmetrically distributed around its nominal  value $\hat{a}_{ij}$. The true distribution of~$\tilde{a}_{ij}$ is unknown and the value of $\tilde{a}_{ij}$ is only known to belong to the \emph{support} $[\hat{a}_{ij}-\overline{a}_{ij},\hat{a}_{ij}+\overline{a}_{ij}]$ of $\tilde{a}_{ij}$, where $\overline{a}_{ij}\geq 0$ is the maximal deviation of the parameter from its nominal (expected)
 value~$\hat{a}_{ij}$. Let~$\mathcal{U}_i$ be the Cartesian product of the supports, i.e. 
\begin{equation}
\mathcal{U}_i=\prod_{j\in [n]}[\hat{a}_{ij}-\overline{a}_{ij},\hat{a}_{ij}+\overline{a}_{ij}],
\label{ui}
\end{equation}
and~$\Gamma_i$ be an integer parameter in $[0,n]$, called \emph{protection level}, which specifies the maximal number of coefficients in the constraint, whose values can be different from their nominal ones.
 Accordingly, define
\begin{equation}
\label{defug}
\mathcal{S}_i=\{\pmb{a}_i=(a_{ij})_{j\in[n]}\in \Rset^n: |\{j: a_{ij}\neq\hat{a}_{ij}\}|\leq \Gamma_i\},
\end{equation}
where $\pmb{a}_i=(a_{ij})_{j\in[n]}$ is a realization (scenario) of  the $i$th  constraint coefficients - a state of the world.
Therefore,
we will consider all scenarios $\pmb{a}_i$ which are in $\mathcal{S}_i\cap \mathcal{U}_i$. Using the robust approach~(\ref{lpr}), we can rewrite the imprecise constraint as 
\begin{equation}
\label{robbert0}
\max_{\pmb{a}_i\in \mathcal{S}_i\cap \mathcal{U}_i} \pmb{a}_i^T\pmb{x}\leq b_i.
\end{equation}
From (\ref{ui}) and (\ref{defug}) and the fact that $\pmb{x}\in \Xset\subset \Rset^n_{+}$ , it follows that (\ref{robbert0})
can be equivalently expressed as
\begin{equation}
\label{robbert}
 \hat{\pmb{a}}_i^T\pmb{x}+\max_{\{N_i\subseteq [n]: |N_i|\leq\Gamma_i\}}\sum_{j\in N_i} \overline{a}_{ij} x_j\leq b_i,
 \end{equation}
where $\hat{\pmb{a}}_i=(\hat{a}_{ij})_{j\in [n]}$ is the vector of  nominal constraint coefficient values.
Making use of the linear programming duality, the inequality~(\ref{robbert}) can be equivalently 
represented as the following system of linear constraints~\cite{BS04} 
(we include the transformation for completeness in Appendix~\ref{dod}):
\begin{equation}
\label{e0}
	\begin{array}{lllll}
		\displaystyle\hat{\pmb{a}}_i^T\pmb{x}+\Gamma_i w_i+\sum_{j\in [n]} p_{ij} \leq b_i\\
		w_i+p_{ij}\geq \overline{a}_{ij}x_j & j\in [n]\\
		w_i\geq 0, p_{ij}\geq 0 & j\in [n],
	\end{array}
\end{equation}
where $w_i$ and $p_{ij}$ are dual variables (see Appendix~\ref{dod}).
Applying~(\ref{e0}) to each constraint $i\in [m]$ we get the following robust counterpart of problem~(\ref{lpf}) that 
is 
consistent with the approach proposed in~\cite{BS04}:
\begin{equation}
\label{bsmod}
	\begin{array}{llll}
			\min & \displaystyle  \pmb{c}^T\pmb{x}\\
			\text{s.t.} &  \multicolumn{2}{l}{\displaystyle\hat{\pmb{a}}_i^T\pmb{x}+\Gamma_i w_i+\sum_{j\in [n]} p_{ij} \leq b_i\;  i\in [m]}\\
				  & w_i+p_{ij}\geq \overline{a}_{ij}x_j &i\in [m], j\in [n]\\
				       &  w_i \geq 0, p_{ij}\geq 0&i\in [m],  j\in[n]\\
 					& \pmb{x}\in \Xset
		\end{array}
\end{equation}
The protection levels $\Gamma_i$, $i\in [n]$, allow  decision makers to control the conservatism of the model
by changing the value of~$\Gamma_i$, from~$0$ to~$n$.
If $\Gamma_i=0$, then only the nominal constraint $\hat{\pmb{a}}_i^T\pmb{x}\leq b_i$ is considered and
the uncertainty is ignored.
 On the other hand, when $\Gamma_i=n$, all the coefficient~$\tilde{a}_{ij}$, $j\in [n]$, 
 can take their  worst-case values. In this case the model becomes the highly conservative  problem~(\ref{lpr}).  It is worth pointing out that the existence of a feasible solution~$\pmb{x}$ to (\ref{bsmod}) depends on~$\Gamma_i$.
 Obviously, by Assumption~\ref{assf},  (\ref{bsmod}) is feasible  if $\Gamma_i=0$ for every $i\in [m]$ and it may be
 infeasible for some larger~$\Gamma_i$, i.e. when 
 the maximum increase in the left hand side of the $i$th constraint (see~(\ref{robbert0}))  for~$\pmb{x}$
 is greater than  the right hand side.
An optimal solution  to~(\ref{bsmod}) for some $\Gamma_i$, $i\in[m]$, prescribed is a robust choice.
Indeed, for this solution we are sure that each constraint~$i$, $i\in[m]$, 
is protected against all scenarios in which at most~$\Gamma_i$
constraint coefficients take values different from their nominal ones - we call such constraints $\Gamma_i$-\emph{protected.}
 However, it is still assumed that a subset of the coefficients will take the largest values in the corresponding supports. The probability of occurrence of the extreme values can be much less than other values within the supports. Model~(\ref{bsmod}) does not take any additional information about the coefficients distributions into account. In the next sections we will extend~(\ref{bsmod}) to the case, in which  possibility distributions for the coefficients are specified.

Under the model of uncertainty assumed in this section, the light robust counterpart of problem~(\ref{lpf}) (see also~(\ref{lplr})) takes the following form~\cite{S14,FM09}:
\begin{equation}
\label{bsmodl}
	\begin{array}{llll}
			\min & \|\pmb{\gamma}\|\\
				\text{s.t.}       &  \multicolumn{2}{l}{\displaystyle\hat{\pmb{a}}_i^T\pmb{x}+\Gamma_i w_i+\sum_{j\in [n]} p_{ij} \leq b_i+\gamma_i\;  i\in [m]}\\
				  & w_i+p_{ij}\geq \overline{a}_{ij}x_j &i\in [m], j\in [n]\\
				  	 &\widehat{\pmb{A}}\pmb{x}\leq \pmb{b}\\
				       & \pmb{c}^T\pmb{x}\leq \hat{c}+\rho_0\\
				       & \gamma_i\geq 0, w_i\geq 0, p_{ij}\geq 0 & i\in [m], j\in [n]\\
 					& \pmb{x}\in \Xset
		\end{array}
\end{equation}
where $w_i$ and $p_{ij}$ are dual variables (see Appendix~\ref{dod}),
$\gamma_i$ is a
slack variable
that  takes  positive value if the $i$th constraint  is violated,
$\hat{c}$ is the optimal objective value of the deterministic counterpart under the nominal scenario $\widehat{\pmb{A}}$, 
$\rho_0\geq 0$ is a fixed tolerance controlling the \emph{price of robustness}, i.e. an increase in the cost of a solution computed
with respect to~$\hat{c}$,
and $\|\cdot\|$ is a given norm
(for instance $\|\cdot\|_1$ or $\|\cdot\|_{\infty}$). The variables $\gamma_i$, $i\in [m]$, in~(\ref{bsmodl}) and
Assumption~\ref{assf} guarantee feasibility and boundedness of~(\ref{bsmodl}). 
Model~(\ref{bsmodl}) is more flexible than~(\ref{bsmod}). It allows us to fix a tradeoff between the robustness of a solution and its price (modeled by the parameter $\rho_0$). However, similarly to model~(\ref{bsmod}), only the information contained in the supports of the uncertain parameters is exploited. 

\section{Possibilistic model of uncertainty}
\label{secpos}

Possibility theory provides a framework of dealing with incomplete information. Its key feature is using two dual set functions, called \emph{possibility} and \emph{necessity measures}. A detailed description of possibility theory can be found in book~\cite{DP88}. We now briefly describe (following~\cite{DD06, DFMP04}) its main components, together with the interpretation assumed in this paper. The primitive object of possibility theory is a \emph{possibility distribution}, which assigns to each element~$u$ in universal set $\Omega$ a degree of possibility $\pi_{\tilde{u}}(u)\in [0,1]$.  Function
$\pi_{\tilde{u}}$ reflects the more or less plausible values of unknown quantity $\tilde{u}$ taking values in $\Omega$. The possibility degree of an event $A\subseteq \Omega$ is then 
$$\Pi(A)=\sup_{u\in A} \pi_{\tilde{u}}(u).$$
Accordingly, the degree of necessity of an event $A\subseteq \Omega$ is
\begin{equation}
\label{necmeasure}
{\rm N}(A)=1-\pi_{\tilde{u}}(\overline{A})=\inf_{u\notin A}(1-\pi_{\tilde{u}}(u)),
\end{equation}
where $\overline{A}$ is the complement of $A$. The necessity measure satisfies the \emph{minitivity axiom}, i.e. 
for any two events $A,B\subseteq \Omega$
\begin{equation}
{\rm N}(A\cap B)=\min\{{\rm N}(A),{\rm N}(B)\}.
\label{mmac}
\end{equation}

There are several interpretations of the possibility and necessity measures. In this paper (see, e.g.,~\cite{DFMP04}) we assume that possibility measure~$\Pi$ encodes the family ${\rm \textbf{P}}(\Pi)$ of probability measures such that ${\rm \textbf{P}}(\Pi)=\{{\rm Pr}\,:\, \forall A \text{ measurable}, {\rm Pr}(A)\leq \Pi(A)\}$ or, equivalently, ${\rm \textbf{P}}(\Pi)=\{{\rm Pr}\,:\, \forall A \text{ measurable}, {\rm Pr}(A)\geq {\rm N}(A)\}$. Hence possibility distribution can be seen as an estimation (upper bound) on the unknown probability distribution, and for each event $A\subseteq \Omega$, ${\rm N}(A)\leq {\rm Pr}(A)\leq \Pi(A)$.
\begin{figure}[ht]
	\centering
	\includegraphics[height=5.3cm]{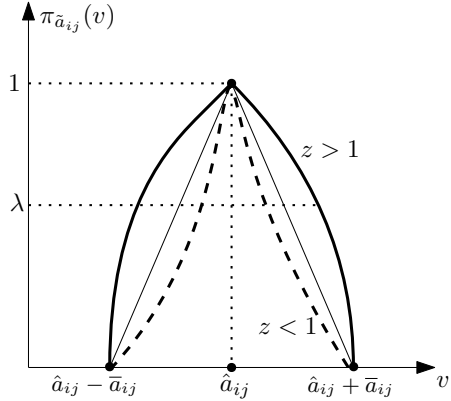}
	\caption{Symmetric fuzzy intervals with $\lambda$-cuts $[\hat{a}_{ij}-\alpha_{ij}(\lambda),\hat{a}_{ij}+\alpha_{ij}(\lambda)]$, where $\alpha_{ij}(\lambda)=\overline{a}_{ij}\cdot(1-\lambda^z)$, $z>0$.}\label{fig1}
\end{figure}

Consider uncertain parameter $\tilde{a}_{ij}$ in matrix $\tilde{\pmb{A}}$. In the approach described in Section~\ref{secrob}, we only know the support $[\hat{a}_{ij}-\overline{a}_{ij},\hat{a}_{ij}+\overline{a}_{ij}]$ of $\tilde{a}_{ij}$. However, in real applications more information about $\tilde{a}_{ij}$ can be provided, which can be utilized to improve the quality of the computed solution. In our model we assume that $\tilde{a}_{ij}$ is a \emph{fuzzy interval}, whose membership function is continuous,  symmetrically distributed around the nominal value $\hat{a}_{ij}$ and with the support equal to $[\hat{a}_{ij}-\overline{a}_{ij},\hat{a}_{ij}+\overline{a}_{ij}]$ (see Figure~\ref{fig1}). The membership function $\pi_{\tilde{a}_{ij}}$ of fuzzy interval~$\tilde{a}_{ij}$ is interpreted as a possibility distribution for $\tilde{a}_{ij}$.

Recall that  the set $\tilde{a}_{ij}^\lambda=\{v \in \Rset: \pi_{\tilde{a}_{ij}}(v)\geq \lambda\}$, $\lambda \in (0,1]$, is called a $\lambda$-\emph{cut} of $\tilde{a}_{ij}$ and contains all values of $\tilde{a}_{ij}$ whose possibility of occurrence is at least $\lambda$. We will assume that $\tilde{a}_{ij}^0$ is the support of $\tilde{a}_{ij}$. The sets $\tilde{a}_{ij}^\lambda=[\hat{a}_{ij}-\alpha_{ij}(\lambda),\hat{a}_{ij}+\alpha_{ij}(\lambda)]$, $\lambda\in [0,1]$, form a nested family of closed intervals with centers equal to the nominal value $\hat{a}_{ij}$. The bound $\alpha_{ij}(\lambda)$ is a continuous, strictly decreasing function in $[0,1]$, such that $\alpha_{ij}(0)=\overline{a}_{ij}$. For example, if $\tilde{a}_{ij}$ is a symmetric triangular fuzzy interval, then $\alpha_{ij}(\lambda)=\overline{a}_{ij}\cdot(1-\lambda)$. One can, however, use also generalized nonlinear functions $\alpha_{ij}(\lambda)=\overline{a}_{ij}\cdot(1-\lambda^z)$, $z>0$, to better reflect the uncertainty (see Figure~\ref{fig1}). Namely, the smaller is the value of $z$ the less uncertainty is associated with $\tilde{a}_{ij}$. For large $z$, $\tilde{a}_{ij}$ tends to a  closed interval. 
Before we proceed, let us state some additional remarks about the model of uncertainty assumed. In the following, 
for simplicity of presentation, we use the same 
value of
$z$ to model the possibility distributions of all imprecise parameters (see Figure~\ref{fig1}). However, the solution method proposed in the next part of the paper can be easily applied to the case in which the values of~$z$ are different. Namely, $\alpha_{ij}(\lambda)=\overline{a}_{ij}\cdot(1-\lambda^{z_{ij}})$ for $z_{ij}>0$. Hence the shapes of the possibility distributions for the parameters can be different, which is reasonable in applications. Also the assumption that the possibility distributions are symmetric, which has been made to be consistent with the interpretation provided in~\cite{BS04}, can be relaxed. We will use this assumption only in simulation tests, performed to compare our approach to the models proposed in~\cite{BS04, FM09, S14}.

Applying~(\ref{necmeasure}) and the continuity of $\pi_{\tilde{a}_{ij}}$ yield
$${\rm N}(\tilde{a}_{ij}^\lambda)=1-\lambda.$$
Hence ${\rm Pr}(\tilde{a}_{ij}^\lambda)\geq 1-\lambda$ and the probability that the value of $\tilde{a}_{ij}$ falls within $\tilde{a}_{ij}^\lambda$ is at least $1-\lambda$.
Let $\pmb{a}_i=(a_{i1},\dots,a_{in})\in \Rset^n$ be scenario describing a realization of $\tilde{\pmb{a}}_i$
(a state of the world)
 in the $i$th imprecise constraint $\tilde{\pmb{a}}_i^T\pmb{x}\leq b_i$.  The degree of possibility that scenario $\pmb{a}_i=(a_{i1},\dots,a_{in})$ will occur is
 provided by 
the following \emph{joint possibility distribution} $\pi_{\tilde{\pmb{a}}_i}$ on
the set of all possible scenarios,  induced by possibility distributions~$\pi_{\tilde{a}_{ij}}$, 
 (see, e.g.,~\cite{DFF03}):
\begin{equation}
\label{pdai}
 \pi_{\tilde{\pmb{a}}_i}(\pmb{a}_i)=\min_{j\in [n]}\pi_{\tilde{a}_{ij}}(a_{ij}).
\end{equation}
We can now compute the set of all scenarios whose possibility of occurrence is at least $\lambda\in (0,1]$ in the following way:
 \begin{eqnarray}
 \mathcal{U}_i^\lambda &  = &    \{\pmb{a}_i\in \Rset^n: \pi_{\tilde{\pmb{a}_i}}(\pmb{a}_i)\geq \lambda\} \nonumber  \\
 	  & = &  \tilde{a}_{i1}^\lambda \times \tilde{a}_{i2}^\lambda\times \dots \times \tilde{a}_{in}^\lambda \label{pdcart1}
\end{eqnarray}
and 
$\mathcal{U}_i^0=\tilde{a}_{i1}^{0}\times \dots \times \tilde{a}_{in}^0$. Now ${\rm N}( \mathcal{U}_i^\lambda)=1-\lambda$, $\lambda\in [0,1]$, so the probability that $\pmb{a}_i$ will fall within $\mathcal{U}_i^\lambda$ is at least $1-\lambda$.

\section{A robust approach to possibilistic optimization problems}
\label{secsfuzzrob}

In this section we generalize the approach proposed in~\cite{BS04} (see Section~\ref{secrob}) to the fuzzy case. We will use the possibilistic interpretation of the uncertain parameters, described in Section~\ref{secpos},
and give a possibilistic counterpart of problem~(\ref{lpf}).

 Consider imprecise constraint $\tilde{\pmb{a}}_i^T\pmb{x}\leq b_i$, in which vector $\tilde{\pmb{a}}_i$ has a possibility distribution described as~(\ref{pdai}). As in Section~\ref{secrob}, we provide a protection level $\Gamma_i$, which is an integer in $[0,n]$ and bounds the number of components in $\tilde{\pmb{a}}_i$ whose realization values are different from their nominal ones. We can now compute the possibility of the event that the constraint will be $\Gamma_i$-protected for a given solution $\pmb{x}\in \Xset$ ($\pmb{x}$ is called $\Gamma_i$-\emph{feasible}):
 \begin{equation}
 \label{pdix}
  \Pi(\pmb{x} \text{  is $\Gamma_i$-\textsc{Feas}})=
        \sup_{\{\pmb{a}_i \in \mathcal{S}_i:\, \pmb{a}_i^T\pmb{x}\leq b_i \}}
        \pi_{\tilde{\pmb{a}}_i}(\pmb{a}_i),
\end{equation} 
where $\mathcal{S}_i$ is defined as~(\ref{defug}). 
 Applying the duality between the possibility and necessity measures gives
  the \emph{degree
 of necessity} that a solution~$\pmb{x}$ is $\Gamma_i$-feasible (see~(\ref{necmeasure})): 
 \begin{equation}
 \mathrm{N}(\pmb{x} \text{  is $\Gamma_i$-\textsc{Feas}})=
        1-\Pi(\pmb{x} \text{  is  not $\Gamma_i$-\textsc{Feas}})
        =1-\sup_{\{\pmb{a}_i \in \mathcal{S}_i:\pmb{a}_i^T\pmb{x} >b_i\}}
        \pi_{\tilde{\pmb{a}}_i}(\pmb{a}_i). \label{ndix}
\end{equation}
Observe that the quantity 
$$\sup_{\{\pmb{a}_i \in \mathcal{S}_i:\pmb{a}_i^T\pmb{x} >b_i\}}
        \pi_{\tilde{\pmb{a}}_i}(\pmb{a}_i)$$
is the possibility of the event that the constraint is not protected, i.e. it can be violated under the assumption that at most $\Gamma_i$ components of $\tilde{\pmb{a}}_i$ are different from their nominal values. Hence $\mathrm{N}(\pmb{x} \text{  is $\Gamma_i$-\textsc{Feas}})\geq 1-\lambda$, $\lambda\in (0,1]$, if and only if for all coefficient scenarios $\pmb{a}_i$ such that  $\pmb{a}_i \in \mathcal{S}_i$ and 
 $\pi_{\tilde{\pmb{a}}_i}(\pmb{a}_i)\geq \lambda$, the inequality $\pmb{a}_i^T\pmb{x}\leq b_i$ holds. 
Using~(\ref{pdcart1}), we get the following proposition:
\begin{prop}
\label{prop1}
For each $\lambda\in [0,1]$, $\mathrm{N}(\pmb{x} \text{  is $\Gamma_i$-\textsc{Feas}})\geq 1-\lambda$ if and only if 
\begin{equation}
\label{e00a}
\max_{\pmb{a}_i\in  \mathcal{S}_i \cap  \mathcal{U}_i^\lambda} \pmb{a}_i^T\pmb{x}\leq b_i.
\end{equation}
\end{prop}

We can now provide the following probabilistic interpretation of our model. If  the inequality $\mathrm{N}(\pmb{x} \text{  is $\Gamma_i$-\textsc{Feas}})\geq 1-\lambda$ holds, then the constraint is $\Gamma_i$-protected with probability at least $1-\lambda$. 
Observe that~(\ref{e00a}) is a parametrized version, with respect to $\lambda$, 
of~(\ref{robbert0}). Hence, it can be replaced with the system of constraints~(\ref{e0}) in which $\overline{a}_{ij}$ is replaced with $\alpha_{ij}(\lambda)=\overline{a}_{ij}\cdot(1-\lambda^z)$
(see also Appendix~\ref{dod}).

Let $\hat{c}$ be the optimal objective value of the deterministic counterpart of problem~(\ref{lpf}) under the nominal scenario $\widehat{\pmb{A}}$ and $\rho_0\geq 0$ be a given tolerance parameter. Consider the crisp constraint
\begin{equation}
\label{objeq}
\pmb{c}^T\pmb{x}\leq \hat{c}+\rho_0,
\end{equation}
which ensures that the cost of solution $\pmb{x}$ must be of some predefined distance from the optimal cost $\hat{c}$. The parameter $\rho_0$ controls the price of robustness of our model (see~\cite{BS04}). Namely, the greater is the value of $\rho_0$ the more relaxed is the optimality of the solution.

Now, given tolerance $\rho_0\geq 0$, we wish to compute a solution, which satisfies all the constraints with the highest necessity degree. Namely, we focus on the following optimization problem:
\begin{equation}
	\label{bnfsp1}
	\textsc{Nec}~\widetilde{\mathcal{P}}:\; \max_{\{\pmb{x}\in \Xset:\; \pmb{c}^T\pmb{x}\leq \hat{c}+\rho_0\} } \mathrm{N}(\wedge^m_{i=1}\; (\pmb{x} \text{  is $\Gamma_i$-\textsc{Feas}})).
\end{equation}
An optimal solution $\pmb{x}^*$ to \textsc{Nec}~$\widetilde{\mathcal{P}}$ is called a \emph{best necessarily feasible  solution}.
Indeed, it is a reasonable choice, because with the highest degree we are sure that it is $\Gamma_i$-feasible for every
$i\in[m]$ and the maximum increase in its cost above~$\hat{c}$ is not greater than~$\rho_0$.
Using the minitivity axiom (see~(\ref{mmac})), we can rewrite~(\ref{bnfsp1}) as follows:
$$
 \max_{\{\pmb{x}\in \Xset:\; \pmb{c}^T\pmb{x}\leq \hat{c}+\rho_0\} } \min_{i\in [m]} \mathrm{N}(\pmb{x}\text{  is $\Gamma_i$-\textsc{Feas}}),
$$
which in turn, by using standard techniques, can be expressed as follows:
\begin{equation}
\label{e02}
	\begin{array}{llll}
		\max  & (1-\lambda) \\
		\text{s.t.} & 	\mathrm{N}(\pmb{x}\text{  is $\Gamma_i$-\textsc{Feas}})\geq 1-\lambda & i\in [m]\\	
		&\pmb{c}^T\pmb{x}\leq \hat{c}+\rho_0\\	
		&0\leq \lambda\leq 1\\	 
		& \pmb{x}\in \Xset
	\end{array}
\end{equation}
By Proposition~\ref{prop1}, we can rewrite~(\ref{e02}) as
\begin{equation}
\label{e04}
	\begin{array}{llll}
		\max  & (1-\lambda) \\
		\text{s.t.}  & \displaystyle \max_{\pmb{a}_i\in  \mathcal{S}_i \cap  \mathcal{U}_i^\lambda} \pmb{a}_i^T\pmb{x}\leq b_i & i\in [m] \\
				& \pmb{c}^T\pmb{x}\leq \hat{c}+\rho_0\\
				&0\leq \lambda\leq 1\\
				 & \pmb{x}\in \Xset
	\end{array}
\end{equation}
Finally, applying~(\ref{e0}), we can represent \textsc{Nec}~$\widetilde{\mathcal{P}}$ as the following mathematical programming problem:
\begin{equation}
\label{bnfsp}
	\begin{array}{llll}
		\max  & (1-\lambda) \\
		\text{s.t.}  &  \multicolumn{2}{l}{\displaystyle \hat{\pmb{a}}_i^T\pmb{x}+\Gamma_i w_i+\sum_{j\in [n]} p_{ij} \leq b_i \;\; i\in [m]}\\
				  & w_i+p_{ij}\geq \alpha_{ij}(\lambda)x_j &i\in [m], j\in [n]\\
				  	&\pmb{c}^T\pmb{x}\leq \hat{c}+\rho_0\\
				        &  w_i \geq 0, p_{ij}\geq 0  &i\in [m],  j\in[n]\\
				       & 0\leq \lambda \leq 1\\
 					& \pmb{x}\in \Xset	
				\end{array}
\end{equation}
where $\alpha_{ij}(\lambda)=\overline{a}_{ij}\cdot(1-\lambda^z)$. 
If $(\pmb{x}^*,\lambda^*)$ is an optimal solution to~(\ref{bnfsp}), then $\pmb{x}^*$ is a best necessarily feasible solution
with $\mathrm{N}(\pmb{x}^*\text{  is $\Gamma_i$-\textsc{Feas}})=1-\lambda^*$.
Note that model~(\ref{bnfsp}) is feasible and bounded by Assumption~\ref{assf} (it is feasible for $\lambda=1$). It is
nonlinear due to the terms $\alpha_{ij}(\lambda) x_{ij}$.
 A method of solving it will be shown in Section~\ref{secsol}.

 \section{A soft robust approach to possibilistic  optimization problems}
 \label{secsoftrob}

\begin{figure}[ht]
	\centering
	\includegraphics[height=5cm]{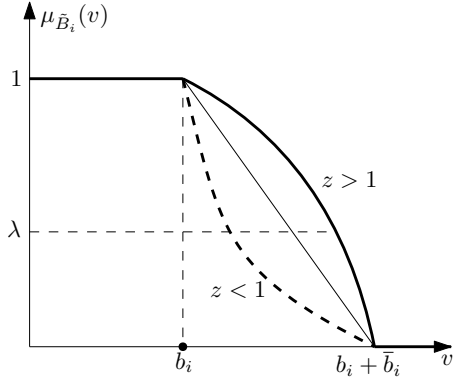}
	\caption{Fuzzy set $\widetilde{B}_i$ with $\mu_{\widetilde{B}_i}^{-1}(\lambda)=b_i+\gamma_i(\lambda)=b_i+\overline{b}_i\cdot(1-\lambda^z)$, $z\geq 0$, representing the right hand side of the $i$th flexible constraint.}\label{fig2}
\end{figure}

In this section we propose a more general and flexible concept for choosing a robust
solution to problem~(\ref{lpf}).  Consider again the uncertain constraint $\tilde{\pmb{a}}_i^T\pmb{x}\leq b_i$, where $\tilde{\pmb{a}}_i$ has a possibility distribution being as in~(\ref{pdai}). Solution~$\pmb{x}$ is feasible for scenario $\pmb{a}_i\in \Rset^n$ if the crisp constraint $\pmb{a}_i^T\pmb{x}\leq b_i$ is satisfied. Following the idea of light robustness~\cite{FM09, S14} (see also~(\ref{lplr})), we relax the concept of feasibility by allowing some violation of the constraint. We assume that $\pmb{x}$ should now satisfy a \emph{flexible constraint} under scenario $\pmb{a}_i$, which is of the form $\pmb{a}^T_i\pmb{x}\widetilde{\leq} \widetilde{B}_i$, where $\widetilde{B}_i$ is a fuzzy set in~$\Rset$ with membership function $\mu_{\widetilde{B}_i}$. The value of $\mu_{\widetilde{B}_i}(\pmb{a}^T_i\pmb{x})$ is the extent to which
$\pmb{a}^T_i\pmb{x}$ satisfies  the flexible constraint. If $\mu_{\widetilde{B}_i}(v)=1$ for $v\leq b_i$ and $\mu_{\widetilde{B}_i}(v)=0$ for $v>b_i$, then 
the flexible constraint reduces to the crisp one.
 In order to model the right hand side of the flexible constraint, we will use fuzzy set $\widetilde{B}_i$, shown in Figure~\ref{fig2}. Namely, $\mu_{\widetilde{B}_i}$ is nonincreasing, $\mu_{\widetilde{B}_i}(v)=1$ for $v\leq b_i$ and $\mu_{\widetilde{B}_i}(v)=0$ for $v\geq b_i+\overline{b}_i$, where $\overline{b}_i\geq 0$ is a parameter denoting the maximal allowed constraint violation.
Let 
$$\mu_{\widetilde{B}_i}^{-1}(\lambda)=\sup\{v : \mu_{\widetilde{B}_i}(v)\geq  \lambda\}, \; \lambda\in (0,1]$$
 be the \emph{pseudoinverse} of 
$\mu_{\widetilde{B}_i}$. We get $\mu_{\widetilde{B}_i}^{-1}(\lambda)=b_i+\gamma_i(\lambda)$, where $\gamma_i(\lambda)$ is nonincreasing function of $\lambda\in [0,1]$ such that $\gamma_i(1)=0$. We will define $\mu_{\widetilde{B}_i}^{-1}(0)=b_i+\gamma_i(0)=b_i+\overline{b}_i$. One can choose, for example, $\gamma_i(\lambda)=\overline{b}_i\cdot(1-\lambda^z)$ for some $z\geq 0$ (see Figure~\ref{fig2}). Notice that the larger is the value of $z$ the larger tolerance for the constraint violation is allowed.
 
We can now compute the possibility of the event that the soft constraint will be $\Gamma_i$-protected for a given solution $\pmb{x}\in \Xset$, i.e. the degree of possibility that $\pmb{x}$ is $\Gamma_i$-\emph{soft feasible}:
  \begin{equation}
  \Pi(\pmb{x}\text{ is $\Gamma_i$-$\widetilde{\textsc{Feas}}$})=
        \sup_{\pmb{a}_i \in \mathcal{S}_i}\min\{\pi_{\tilde{\pmb{a}}_i}(\pmb{a}_i), \mu_{\widetilde{B}_i}(\pmb{a}_i^T\pmb{x})\}.\label{pdix1}\\
\end{equation} 
 Notice that in~(\ref{pdix1}) we jointly consider the 
  uncertainty (induced by the uncertain coefficients in~$\tilde{\pmb{a}}_i$)
and  flexibility
 of the $i$th constraint (see~\cite{DFF03}). 
 Accordingly, the degree
 of necessity that a solution~$\pmb{x}$ is $\Gamma_i$-soft feasible
 is defined as follows:
 \begin{align}
 &\mathrm{N}(\pmb{x}\text{ is $\Gamma_i$-$\widetilde{\textsc{Feas}}$})=
        1-\Pi(\pmb{x}\text{ is  not $\Gamma_i$-$\widetilde{\textsc{Feas}}$}) \label{sndix}\\
       & =1-\sup_{\pmb{a}_i \in \mathcal{S}_i}\min\{\pi_{\tilde{\pmb{a}}_i}(\pmb{a}_i),
       1-\mu_{\widetilde{B}_i}(\pmb{a}^T_i\pmb{x})\}.\nonumber
\end{align}
Thus $\mathrm{N}(\pmb{x} \text{ is $\Gamma_i$-$\widetilde{\textsc{Feas}}$})\geq 1-\lambda$, $\lambda\in [0,1]$,
if and only if for all scenarios~$\pmb{a}_i$
such that $\pmb{a}_i \in \mathcal{S}_i$ and $\pi_{\tilde{\pmb{a}}_i}(\pmb{a}_i)\geq \lambda$, the inequality $\mu_{\widetilde{B}_i}(\pmb{a}^T_i\pmb{x})\geq 1-\lambda$ holds. This inequality is equivalent to $\pmb{a}_i^T\pmb{x}\leq \mu_{\widetilde{B}_i}^{-1}(1-\lambda)=b_i+\gamma_i(1-\lambda).$ Hence, (\ref{pdcart1}) leads to the following proposition:
\begin{prop}
\label{prop4}
For each $\lambda\in [0,1]$, $\mathrm{N}(\pmb{x} \text{ is $\Gamma_i$-$\widetilde{\textsc{Feas}}$})\geq 1-\lambda$ if and only if 
\begin{equation}
\label{e00}
\max_{\pmb{a}_i\in  \mathcal{S}_i \cap  \mathcal{U}_i^\lambda} \pmb{a}_i^T\pmb{x}\leq b_i+\gamma_i(1-\lambda),
\end{equation}
where $\gamma_i(1-\lambda)=\overline{b}_i\cdot(1-(1-\lambda)^z)$.
\end{prop}

 We can now provide the following probabilistic interpretation of our model. If  the inequality $\mathrm{N}(\pmb{x} \text{ is $\Gamma_i$-$\widetilde{\textsc{Feas}}$})\geq 1-\lambda$ holds, then the $i$th constraint is $\Gamma_i$-protected with the tolerance $\gamma_i(1-\lambda)$, with probability at least $1-\lambda$. 
  
 In the approach described in Section~\ref{secsfuzzrob} we required that  $\pmb{c}^T\pmb{x}\leq \hat{c}+\rho_0$, where $\hat{c}$ is the optimal objective value of the deterministic counterpart under the nominal scenario $\widehat{\pmb{A}}$ and $\rho_0\geq 0$ is the assumed tolerance.  We can replace this crisp constraint with a flexible constraint of the form $\pmb{c}^T\pmb{x} \widetilde{\leq} \widetilde{C}$, where $\widetilde{C}$ is a fuzzy set shown in Figure~\ref{fig2}, with the pseudoinverse $\mu^{-1}_{\widetilde{C}}(\lambda)=\hat{c}+\zeta(\lambda)=\hat{c}+\rho_0\cdot(1-\lambda^z)$, where the interpretation of $\hat{c}$ and $\rho_0$ is the same as in Section~\ref{secsfuzzrob}.  Now, $\mu_{\widetilde{C}}(\pmb{c}^T\pmb{x})$ expresses a preference (satisfaction) about the deviation of $\pmb{c}^T\pmb{x}$ from $\hat{c}$ (less deviations are more preferred).
  We can define the necessity degree that the flexible constraint $\hat{\pmb{c}}^T\pmb{x} \widetilde{\leq} \widetilde{C}$ is satisfied as follows:
\begin{equation}
\label{necc0}
	\mathrm{N}(\pmb{c}^T\pmb{x} \widetilde{\leq} \widetilde{C})=
        1-\sup_{\{c:\; \pmb{c}^T\pmb{x}>c\}} \mu_{\widetilde{C}}(c),
\end{equation}
The following proposition is analogous to Proposition~\ref{prop4}:
\begin{prop}
\label{prop5}
For each $\lambda\in [0,1]$, $\mathrm{N}(\pmb{c}^T\pmb{x} \widetilde{\leq} \widetilde{C})\geq 1-\lambda$ if and only if 
\begin{equation}
\label{e003}
\pmb{c}^T\pmb{x}\leq \hat{c}+\zeta(1-\lambda),
\end{equation}
where $\zeta(1-\lambda)=\rho_0\cdot(1-(1-\lambda)^z)$.
\end{prop}
Note that we can control the flexibility of the constraint $\pmb{c}^T\pmb{x} \widetilde{\leq} \widetilde{C}$ by changing the parameter~$z$. If $z=0$, then  the computed solution must be optimal under the nominal scenario. On the other hand, if $z>0$ is large, then the constraint tends to the crisp constraint $\pmb{c}^T\pmb{x}\leq \hat{c}+\rho_0$, which was used in the model discussed in Section~\ref{secsfuzzrob}.

We can now extend model~(\ref{bnfsp1}) by considering the following optimization problem:
\begin{equation}
	\label{bnfsps}
	\textsc{Soft-Nec}~\widetilde{\mathcal{P}}: \max_{\pmb{x}\in \Xset} \mathrm{N}(\wedge^m_{i=1}(\pmb{x} \text{  is } \Gamma_i-\widetilde{\textsc{Feas}}) \wedge (\pmb{c}^T\pmb{x} \widetilde{\leq} \widetilde{C})).
\end{equation}
An optimal solution~$\pmb{x}^*$ to~(\ref{bnfsps}) is
called a \emph{best necessary soft feasible}. 
Such a solution maximizes the necessity degree that it is
$\Gamma_i$-soft feasible for every
$i\in[m]$
and
 its cost  $\pmb{c}^T\pmb{x}^*$ falls within fuzzy cost~$\widetilde{C}$.
Using the minitivity axiom, Proposition~\ref{prop4} and~\ref{prop5}, and applying the same reasoning as in Section~\ref{secsfuzzrob}, we can represent $\textsc{Soft-Nec}~\widetilde{\mathcal{P}}$ as follows:
\begin{equation}
\label{bnsmod}
 \begin{array}{llll}
                    \max      &(1-\lambda)  \\
       \text{s.t.}                    & \multicolumn{2}{l}{\displaystyle \hat{\pmb{a}}^T_{i} \pmb{x}+\Gamma_iw_i+\sum_{j\in [n]}p_{ij}\leq b_i +\gamma_i(1-\lambda)\;  i\in [m]}  \\
                                      &w_i+p_{ij}\geq \alpha_{ij}(\lambda)x_j & i\in [m],  j\in [n] \\
                                      & \pmb{c}^T\pmb{x}\leq \hat{c}+\zeta(1-\lambda)\\
				       & w_i, p_{ij}\geq 0 & i\in [m], j\in [n]\\ 
				       &0\leq \lambda\leq 1\\
                                       &\pmb{x}\in \Xset
 \end{array}
 \end{equation}
 where $\alpha_{ij}(\lambda)=\overline{a}_{ij}\cdot(1-\lambda^z)$, $\gamma_i(1-\lambda)=\overline{b}_i\cdot(1-(1-\lambda)^z)$, $\zeta(1-\lambda)=\rho_0\cdot(1-(1-\lambda)^z)$.
 If $(\pmb{x}^*,\lambda^*)$ is an optimal solution to~(\ref{bnsmod}), then $\pmb{x}^*$ is a best necessarily soft  feasible solution
with  $\mathrm{N}(\wedge^m_{i=1}(\pmb{x}^* \text{  is } \Gamma_i-\widetilde{\textsc{Feas}}) \wedge 
(\pmb{c}^T\pmb{x}^* \widetilde{\leq} \widetilde{C}))=1-\lambda^*$. Such a solution exists, since by Assumption~\ref{assf}
model~(\ref{bnsmod}) is
feasible and bounded. Note that it is  nonlinear.
We will show a method of solving~(\ref{bnsmod}) in Section~\ref{secsol}.
 One can also optionally add to~(\ref{bnsmod}),  along the  lines of~\cite{FM09,S14},
 the crisp constraints
\begin{equation}
\label{cfn}
	\widehat{\pmb{A}}\pmb{x}\leq \pmb{b}
\end{equation}
ensuring the feasibility of the solution in the nominal scenario.

\subsection{Illustrative example}
\label{secexamp}

Consider the following uncertain problem~(\ref{lpf}):
\begin{equation}
\label{examp}	
	\begin{array}{lll}
			\min &  -4x_1-3x_2-2x_3-x_4 \\
			\text{s.t.} & \braket{0,7} x_1+\braket{1,5} x_2+\braket{2,4}x_3+\braket{3,2}x_4\leq 6\\
				       & 0\leq x_i\leq 1 \;\; i\in [4],
	\end{array}
\end{equation}
where $\braket{\hat{a}_j, \overline{a}_{j}}$, $j\in [n]$, are symmetric triangular fuzzy intervals with supports $[\hat{a}_{j}-\overline{a}_{j},\hat{a}_{j}+\overline{a}_{j}]$, respectively. An optimal solution to the nominal problem, i.e. the one with $\hat{\pmb{a}}=(0,1,2,3)$,  is $(1,1,1,1)$ with $\hat{c}=-10$. The robustness of this solution is weak as the constraint violation is highly probable (it is worth pointing out that an increase of any coefficient above its nominal value results in 
solution infeasibility.)
 Let us fix the protection level $\Gamma_1=2$, so the values of at most two coefficients in the constraint can differ from their nominal ones.  
If we use the robust model~(\ref{bsmod}) for the supports of the fuzzy intervals, namely $([-7,7], [-4,6],[-2,6],[1,5])$, then we get an optimal solution $\pmb{x}'=(0.325,0.437,0.547,0)$ with the objective value~$-3.71$.  Notice that the possibilistic information for $\tilde{\pmb{a}}$ is not taken into account. This solution is more protected against the constraint violation, but  one can observe a large deterioration ($|\frac{\pmb{c}^T\pmb{x}'-\hat{c}}{\hat{c}}|\cdot 100\%=62.9\%$) in the optimal objective value, so $\pmb{x}'$ has a large price of robustness. 

\begin{figure}[ht]
	\centering
	\includegraphics[height=5.5cm]{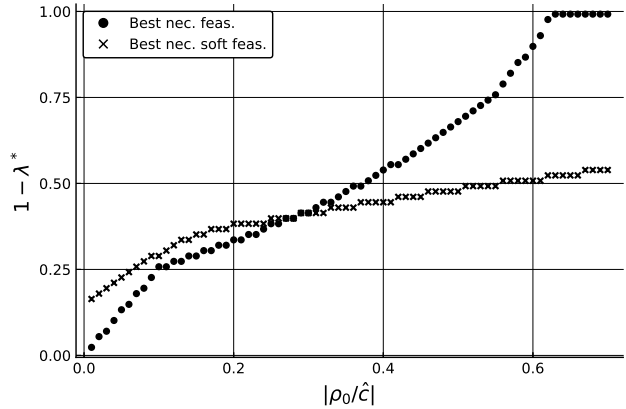}
	\caption{The optimal objective values of~(\ref{bnfsp}) and~(\ref{bnsmod}), depending on the ratio $|\rho_0/\hat{c}|$.}\label{fig3a}
\end{figure}

Let us now investigate the effect of taking the complete possibilistic information about $\tilde{\pmb{a}}$ into account.
 We compute a best necessarily feasible solution to~(\ref{examp}) by solving the corresponding model~(\ref{bnfsp}) with $\Gamma_1=2$. We can now control the price of robustness of the solution by changing the tolerance $\rho_0$, used in the constraint $\pmb{c}^T\pmb{x}\leq \hat{c}+\rho_0$. In Figure~\ref{fig3a}, the optimal objective value of~(\ref{bnfsp}), i.e. the degree of $\Gamma_1$-feasibility, depending on the ratio $|\rho_0/\hat{c}|$ is shown. If $\rho_0=0$, then we require that the solution computed must be optimal for the nominal scenario. In this case, the best necessarily feasible solution is $\pmb{x}=(1,1,1,1)$ and its degree of necessary $\Gamma_1$-feasibility is~0. On the other hand, if we fix $\rho_0\geq 6.29$ (the ratio $|\rho_0/\hat{c}|\geq 0.629$), the best necessarily feasible solution computed is the same as the optimal robust solution to~(\ref{bsmod}) for $\Gamma_1=2$ (recall that the optimal objective value of~(\ref{bsmod}) is $-10+0.629=-3.71$).  The degree of  necessary $\Gamma_1$-feasibility of this solution equals~1.
 It can be reasonable to choose some intermediate value of $\rho_0\in[0, 6.29]$. For example, if $\rho_0=3$ (the ratio $|\rho_0/\hat{c}|=0.3$), then we get solution $\pmb{x}^*=(1, 0.6, 0.6,0)$ with the degree of necessary $\Gamma_1$-feasibility equal to 0.44. 
\begin{figure}[ht]
	\centering
	\includegraphics[height=5.5cm]{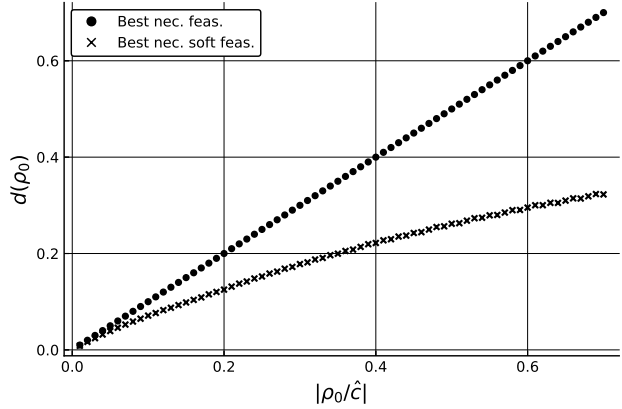}
	\caption{The ratio $d(\rho_0)=|(\pmb{c}^T\pmb{x}^*-\hat{c})/\hat{c}|$, where $\pmb{x}^*$ is an optimal solution to
	 ~(\ref{bnfsp}) or~(\ref{bnsmod}), depending on the ratio $|\rho_0/\hat{c}|$.}\label{fig3b}
\end{figure}

Let us now compute a best necessarily soft feasible solution to~(\ref{examp}) by solving~(\ref{bnsmod}). Assume that the maximum accepted magnitude of the constraint violation equals~$\overline{b}_1=2$, i.e. it is at most~$33\%$ of its nominal value equal to~6. The crisp right hand side in~(\ref{examp}) is thus replaced with fuzzy set $\widetilde{B}_1$ with $\mu^{-1}_{\widetilde{B}_1}(\lambda)=6+2(1-\lambda)$. We also replace the crisp constraint $\pmb{c}^T\pmb{x}\leq \hat{c}+\rho_0$ with the flexible constraint $\pmb{c}^T\pmb{x}\widetilde{\leq}\widetilde{C}$, where $\widetilde{C}$ is a fuzzy set with the pseudoinverse $\hat{c}+\rho_0\cdot(1-\lambda)$. As in the previous model,  $\hat{c}=-10$ and $\rho_0$ is a parameter denoting the maximum accepted tolerance,  controlling the price of robustness of the solution computed.

Let us first investigate the deterioration of the objective function for various $\rho_0$ (see Figure~\ref{fig3b}). Let $\pmb{x}^*$ be an optimal solution to~(\ref{bnfsp}) or~(\ref{bnsmod}) for a fixed $\rho_0$ and consider the ratio $d(\rho_0)=|(\pmb{c}^T\pmb{x}^*-\hat{c})/\hat{c}|$. Observe that for~(\ref{bnfsp}) the ratio $d(\rho_0)$ increases linearly with $|\rho_0/\hat{c}|$. This is due to the constraint $\pmb{c}^T\pmb{x}\leq \hat{c}+\rho_0$, which is tight at $\pmb{x}^*$.
 Different behavior can be observed if $\pmb{x}^*$ is an optimal solution to~(\ref{bnsmod}). In general, the ratio $d(\rho_0)$ can be smaller, due to the constraint $\pmb{c}^T\pmb{x}\leq \hat{c}+\zeta(1-\lambda)=\hat{c}+\rho_0\cdot\lambda$, which is tight at $\pmb{x}^*$ and $\lambda^*\in [0,1]$. Hence, model~(\ref{bnsmod}) returns solutions with smaller price of robustness.
 
  In Figure~\ref{fig3a} the optimal objective values of~(\ref{bnfsp}) and~(\ref{bnsmod}) are compared. For smaller ratios $|\rho_0/\hat{c}|$ the objective value of~(\ref{bnsmod}) is greater. This is the effect of relaxation of the constraint which dominates the preference imposed on the objective value. The situation reverses for larger ratios $|\rho_0/\hat{c}|$, where the preference about the objective value is relaxed. Then a~solution computed has a smaller price of robustness but also is less protected against the constraint violation.
  
  In order to test the quality of the obtained solutions, one can perform a simulation, i.e. test the feasibility of the model for a sample of scenarios drawn according to the joint possibility distribution for $\tilde{\pmb{a}}_i$. Such a simulation for larger instances will be done in Section~\ref{secexper}.

\section{Treating the uncertain objective function}
\label{secfobj}

 In this section we will show how the model discussed in Section~\ref{secsoftrob}  can be extended to handle the uncertainty in the objective function into account. Suppose that the vector of objective function coefficients in~(\ref{lpf}), denoted now 
by $\tilde{\pmb{c}}$, is imprecise.
 Many approaches have been proposed in the literature to deal with imprecise objective function $\tilde{\pmb{c}}^T\pmb{x}$. In the fuzzy setting, the problem is often reduced to minimizing $r(\tilde{\pmb{c}}^T\pmb{x})$, where $r$ is a real-valued ranking function~\cite{CV00,CZ00a,FR82,PZT11}. In another approach, a fuzzy goal $\tilde{g}$ is associated with the imprecise objective function and one can maximize ${\rm N}(\tilde{\pmb{c}}^T\pmb{x}\leq\widetilde{G})$, which is interpreted as the necessity degree of achieving the goal $\widetilde{G}$.  This concept can be softened~\cite{IS98,I16} by maximizing ${\rm N}(\tilde{z}(\pmb{x})\leq\tilde{g})$, where $\tilde{z}(\pmb{x})$ is a fuzzy set whose membership function describes a possibility distribution of the maximum regret of $\pmb{x}$ (the maximum distance to the optimality of $\pmb{x}$).
 
 In this section we will propose a method of dealing with uncertain vector $\tilde{\pmb{c}}$, which is analogous to the concept described in the previous sections for the uncertain constraints.  We will apply an approach, commonly used in robust and stochastic optimization (see, e.g.,~\cite{BS04}), 
 which consists in representing the imprecise objective function as 
 imprecise constraint $\tilde{\pmb{c}}^T\pmb{x}-x_0\leq 0$
 and minimizing~$x_0$,
 where $x_0$ is 
 an additional variable  that reflects possible realizations of  objective function values.
 Therefore, we now study the following problem:
 \begin{equation}
 \begin{array}{lll}
\min &\displaystyle x_0\\
  \text{s.t.}	      & \tilde{\pmb{c}}^T\pmb{x}-x_0\leq 0  \\
  		& \widetilde{\pmb{A}}\pmb{x}\leq \pmb{b} \\
 		& \pmb{x}\in \Xset
  \end{array}
 \label{lpr1}
\end{equation}
Observe that~(\ref{lpr1}) has deterministic objective function and one additional imprecise constraint of the form $\tilde{\pmb{c}}^T\pmb{x}-x_0\leq 0$. Hence, it is of the form~(\ref{lpf}) and for deterministic $\tilde{\pmb{c}}$ it is equivalent to~(\ref{lpf}). We can now treat this new constraint just in the same way as the remaining imprecise constraints.
 
In order to define $\tilde{\pmb{c}}$, we will use the possibilistic model of uncertainty, described in Section~\ref{secpos}. Namely, $\tilde{c}_j$, $j\in [n]$, are fuzzy intervals with membership functions $\pi_{\tilde{c}_j}$, symmetrically distributed around the nominal values $\hat{c}_j$ and with the supports $[\hat{c}_j-\overline{c}_j, \hat{c}_j+\overline{c}_j]$. We will use $\tilde{c}_j^\lambda=[\hat{c}_j-\beta_j(\lambda),\hat{c}_j+\beta_j(\lambda)]$ to denote the $\lambda$-cut of $\tilde{c}_j$, where $\beta_j(\lambda)=\overline{c}_j(1-\lambda^z)$ for a fixed $z>0$ (see Figure~\ref{fig1}).
 If $\pmb{c}\in \Rset^n$ is scenario describing a realization of the uncertain objective function coefficients, then after applying the same reasoning as previously (see~(\ref{pdai})), we can compute
$$\pi_{\tilde{\pmb{c}}}(\pmb{c})=\min_{j\in [n]} \pi_{\tilde{c}_j}(c_j).$$
Then
 $\mathcal{U}_0^\lambda= \{\pmb{c}\in \Rset^n: \pi_{\tilde{\pmb{c}}}(\pmb{c})\geq \lambda\} = \tilde{c}_{1}^\lambda \times \tilde{c}_{2}^\lambda\times \dots \times \tilde{c}_{n}^\lambda$
and 
$\mathcal{U}_0^0=\tilde{c}_{1}^{0}\times \dots \times \tilde{c}_{n}^0$. Now  ${\rm N}( \mathcal{U}_0^\lambda)=1-\lambda$, $\lambda\in [0,1]$, so the probability that $\pmb{c}$ will fall within $\mathcal{U}_0^\lambda$ is at least $1-\lambda$.

Let us define a protection level $\Gamma_0$, being an integer in $[0,n]$. Then 
$$\mathcal{S}_0=\{(c_{j})_{j\in[n]}\in \Rset^n: |\{j: c_{j}\neq\hat{c}_{j}\}|\leq \Gamma_0\}.$$
Let us introduce fuzzy set $\widetilde{B}_0$ (see Figure~\ref{fig2}) with  pseudoinverse $\mu_{\widetilde{B}_0}^{-1}(\lambda)=\gamma_0(\lambda)=\overline{b}_0\cdot(1-\lambda^z)$. Accordingly, we can define
\begin{align}
 \mathrm{N}((x_0,\pmb{x})\text{ is $\Gamma_0$-$\widetilde{\textsc{Feas}}$})&=
        1-\Pi((x_0,\pmb{x})\text{ is  not $\Gamma_0$-$\widetilde{\textsc{Feas}}$}) \nonumber\\
       & =1-\sup_{\pmb{c} \in \mathcal{S}_0}\min\{\pi_{\tilde{\pmb{c}}}(\pmb{c}),
       1-\mu_{\widetilde{B}_0}(\pmb{c}^T\pmb{x}-x_0)\}.\label{sndix1}
\end{align}
The following proposition is analogous to Proposition~\ref{prop4}:
\begin{prop}
\label{prop6}
For each $\lambda\in [0,1]$, $\mathrm{N}((x_0,\pmb{x}) \text{ is $\Gamma_0$-$\widetilde{\textsc{Feas}}$})\geq 1-\lambda$ if and only if 
\begin{equation}
\label{e007}
\max_{\pmb{c}\in  \mathcal{S}_0\cap  \mathcal{U}_0^\lambda} \pmb{c}^T\pmb{x}-x_0\leq \gamma_0(1-\lambda),
\end{equation}
where $\gamma_0(1-\lambda)=\overline{b}_0\cdot(1-(1-\lambda)^z)$.
\end{prop}

Let $\hat{c}$ be the optimal objective value of the deterministic counterpart of~(\ref{lpr1}) under the nominal scenario $(\widehat{\pmb{A}}, \hat{\pmb{c}})$. The flexible constraint $\pmb{c}^T\pmb{x}\widetilde{\leq}\widetilde{C}$, considered in Section~\ref{secsoftrob}, becomes then $x_0\widetilde{\leq} \tilde{C}$, where $\widetilde{C}$ is defined in the same way as in Section~\ref{secsoftrob}. We can now extend \textsc{Soft-Nec}~$\widetilde{P}$ (see~(\ref{bnfsps})), by using  the necessity degree of conjunction of the  events:
$$\wedge^m_{i=1}(\pmb{x} \text{  is } \Gamma_i-\widetilde{\textsc{Feas}})\wedge ((\pmb{x},x_0) \text{  is } \Gamma_0-\widetilde{\textsc{Feas}}) \wedge 
(x_0 \widetilde{\leq} \widetilde{C}),$$
to the following  optimization problem:
\begin{equation}
\label{ebnfsps}
\textsc{Soft-Nec}~\widetilde{\mathcal{P}}: \max_{\pmb{x}\in \Xset} \mathrm{N}(\wedge^m_{i=1}(\pmb{x} \text{  is } \Gamma_i-\widetilde{\textsc{Feas}})\wedge ((\pmb{x},x_0) \text{  is } \Gamma_0-\widetilde{\textsc{Feas}}) \wedge 
(x_0 \widetilde{\leq} \widetilde{C})).
\end{equation}
Taking Proposition~\ref{prop6} into account and applying the same reasoning as in Section~\ref{secsoftrob}, we can represent \textsc{Soft-Nec}~$\mathcal{\widetilde{P}}$ as the following mathematical programming problem:
\begin{equation}
\label{bnsmoduc}
 \begin{array}{llll}
                         \max  &(1-\lambda) \\
                    \text{s.t.}      &  \multicolumn{2}{l}{\displaystyle \hat{\pmb{c}}^T \pmb{x}+\Gamma_0w_0+\sum_{j\in [n]}q_j-x_0\leq \gamma_0(1-\lambda)} \\
                                      &w_0+q_{j}\geq \beta_j(\lambda)x_j  &   j\in [n] \\
                           & \multicolumn{2}{l}{\displaystyle \hat{\pmb{a}}^T_{i} \pmb{x}+\Gamma_iw_i+\sum_{j\in [n]}p_{ij}\leq b_i +\gamma_i(1-\lambda)\;  i\in [m]}  \\
                                      &w_i+p_{ij}\geq \alpha_{ij}(\lambda)x_j & i\in [m],  j\in [n] \\
                                       & x_0\leq \hat{c}+\zeta(1-\lambda)\\
                                       & w_i\geq 0 & i\in [m]\cup\{0\} \\
                                        & q_j \geq 0, p_{ij}\geq 0 & i\in [m], j\in[n]\\
                                        &0\leq \lambda\leq 1\\
                                       &\pmb{x}\in \Xset
 \end{array}
 \end{equation}
 Observe that  the variable~$x_0$ can be eliminated from~(\ref{bnsmoduc}), which yields:
\begin{equation}
\label{bnsmoduc1}
 \begin{array}{llll}
                         \max  &(1-\lambda) \\
                    \text{s.t.}      &  \multicolumn{2}{l}{\displaystyle \hat{\pmb{c}}^T \pmb{x}+\Gamma_0w_0+\sum_{j\in [n]}q_j\leq \hat{c}+\zeta(1-\lambda)+ \gamma_0(1-\lambda)} \\
                                      &w_0+q_{j}\geq \beta_j(\lambda)x_j  &   j\in [n] \\
                           & \multicolumn{2}{l}{\displaystyle \hat{\pmb{a}}^T_{i} \pmb{x}+\Gamma_iw_i+\sum_{j\in [n]}p_{ij}\leq b_i +\gamma_i(1-\lambda)\;  i\in [m]}  \\
                                      &w_i+p_{ij}\geq \alpha_{ij}(\lambda)x_j & i\in [m],  j\in [n] \\
                                      & w_i\geq 0 & i\in [m]\cup\{0\}\\
                                        &  q_j \geq 0, p_{ij}\geq 0 & i\in[m], j\in[n]\\
                                        &0\leq \lambda\leq 1\\
                                       &\pmb{x}\in \Xset
 \end{array}
 \end{equation}
 where $\alpha_{ij}(\lambda)=\overline{a}_{ij}\cdot(1-\lambda^z)$, $\beta_j(\lambda)=\overline{c}_j\cdot(1-\lambda^z)$, $\gamma_i(1-\lambda)=\overline{b}_i\cdot(1-(1-\lambda)^z)$ and $\zeta(1-\lambda)=\rho_0\cdot(1-(1-\lambda)^z)$.
 If $(\pmb{x}^*,\lambda^*)$ is an optimal solution to~(\ref{bnsmoduc1}), then $\pmb{x}^*$ is a best necessarily soft  feasible solution
with  $\mathrm{N}(\wedge^m_{i=1}(\pmb{x}^* \text{  is } \Gamma_i-\widetilde{\textsc{Feas}})\wedge ((\pmb{x}^*,x_0) \text{  is } \Gamma_0-\widetilde{\textsc{Feas}}) \wedge 
(x_0 \widetilde{\leq} \widetilde{C}))=1-\lambda^*$.  
Note that by Assumption~\ref{assf}
model~(\ref{bnsmoduc1}) is
feasible and bounded. It is nonlinear and a method of solving it will be shown in Section~\ref{secsol}.
Model~(\ref{bnsmoduc1}) generalizes~(\ref{bnfsp}) and~(\ref{bnsmod}). Indeed, if there  is no uncertainty in the objective, then $\pmb{c}=\hat{\pmb{c}}$, $\overline{b}_0=0$, and $\overline{c}_j=0$ for each $j\in [n]$. Then the first two constraints of~(\ref{bnsmoduc1}) reduce to $\pmb{c}^T\pmb{x}\leq \hat{c}+ \zeta(1-\lambda)$, which yields~(\ref{bnsmod}). Fixing further large $z$ in $\zeta(1-\lambda)=\rho_0\cdot(1-(1-\lambda)^z)$ and $\overline{b}_i=0$ for all $i\in [m]$ leads to~(\ref{bnfsp}).
 
 \section{Solving the problem}
 \label{secsol}

The problems arising in practice are often of large-scale. It is thus important to construct efficient algorithms to solve them. In this section we show that the complexity of solving the uncertain problem under consideration is essentially the same as the complexity of solving its deterministic counterpart.
For a brief introduction to computational complexity theory,  we refer the reader to~\cite[Chapter 34]{CO09}.

Let us focus on solving $\textsc{Soft-Nec}~\widetilde{\mathcal{P}}$  (see~(\ref{ebnfsps})). We will study the most general model~(\ref{bnsmoduc1}), in which an uncertain objective function is taken into account. For a fixed value of $\lambda\in [0,1]$, all the constraints in~(\ref{bnsmoduc1}) (possibly, except for the ones describing $\pmb{x}\in \Xset$) become linear. Let $\Xset^\lambda\subseteq \Xset$ be the set of feasible solutions to~(\ref{bnsmoduc1}) for a fixed value of $\lambda\in [0,1]$. Since all the functions $\alpha_{ij}(\lambda)$, $\beta_j(\lambda)$, $\gamma_i(\lambda)$, $\zeta(\lambda)$ are nonincreasing,  we get $\Xset^{\lambda_1}\subseteq \Xset^{\lambda_2}$ if $\lambda_1\leq \lambda_2$. Consequently,~(\ref{bnsmoduc1}) can be solved by computing the smallest value $\lambda_{\min}\in [0,1]$ for which $\Xset^{\lambda_{\min}}$ is nonempty. This can be done by applying a binary search in the interval $[0,1]$ (see Algorithm~\ref{algbnfs}).
\begin{algorithm}
\begin{small}
\caption{\small Solving \textsc{Soft-Nec} $\widetilde{\mathcal{P}}$ with accuracy $\epsilon>0$
}\label{algbnfs}
   	$\overline{\lambda}\leftarrow 1$, $\underline{\lambda}\leftarrow 0$\;
	$\hat{c}\leftarrow\hat{\pmb{c}}^T\pmb{x}^*=
	\min\{\hat{\pmb{c}}^T\pmb{x} : \widehat{\pmb{A}}\pmb{x}\leq \pmb{b}, \pmb{x}\in \Xset\}$  \label{algbnfs5a} \;
	\While{$|\overline{\lambda}-\underline{\lambda}|> \epsilon$}
	{
	$\lambda\leftarrow \underline{\lambda}+(\overline{\lambda}-\underline{\lambda})/2$\;
	\If{\text{there exists $\pmb{x}$ feasible to~(\ref{bnsmoduc1})	
	 for $\lambda$}  \label{algbnfs5}} 
         {
		$\pmb{x}^*\leftarrow \pmb{x}$,
   		$\overline{\lambda}\leftarrow \lambda$	    
         }
	\lElse{$\underline{\lambda}\leftarrow \lambda$}
	}
	\Return{$\pmb{x}^*$,
	$1-\overline{\lambda}$}\;
        \tcp{A best necessarily soft feasible  solution~$\pmb{x}^*$}
	\tcp{$\mathrm{N}(\wedge^m_{i=1}(\pmb{x}^* \text{  is } \Gamma_i-\widetilde{\textsc{Feas}})\wedge ((\pmb{x}^*,x_0) \text{  is } \Gamma_0-\widetilde{\textsc{Feas}}) \wedge (x_0 \widetilde{\leq} \widetilde{C}))=1-\overline{\lambda}
$}
	\end{small}	
\end{algorithm}

The running time of Algorithm~\ref{algbnfs} depends of the complexity of the problem which must be solved in 
Steps~\ref{algbnfs5a} and~\ref{algbnfs5}, i.e. checking the feasibility of~(\ref{bnsmoduc1}) for a fixed $\lambda\in [0,1]$.
In Step~\ref{algbnfs5a} the feasibility of~(\ref{bnsmoduc1}) is implicitly checked for $\lambda=1$.
Indeed, it is easily seen
that this task can be reduced to solving  the deterministic counterpart of problem~(\ref{lpf}) under the nominal scenario $(\widehat{\pmb{A}}, \hat{\pmb{c}})$, since
such solution~$\pmb{x}^*$ computed, whose existence  follows from  Assumption~\ref{assf},  is always feasible  to~(\ref{bnsmoduc1})
for $\lambda=1$.
Thus the computational complexity of Steps~\ref{algbnfs5a} and~\ref{algbnfs5}
 depends on the structure of the set $\Xset$. If the feasibility can be checked in $T(|I|)$ time, where $|I|$ is the size of~(\ref{bnsmoduc1}), then Algorithm~\ref{algbnfs} runs in $O(\lceil\log \epsilon^{-1} \rceil T(|I|))$ time, because the feasibility must be tested at most $\lceil\log \epsilon^{-1} \rceil+1$ times. If $T(|I|)$ is polynomial in size~$|I|$, then Algorithm~\ref{algbnfs} runs in polynomial time and \textsc{Soft-Nec}~$\widetilde{\mathcal{P}}$ can be solved in polynomial time with a fixed accuracy $\epsilon>0$. In the next section we will identify some important special cases of problem~(\ref{lpf}) for which this is the case.

\subsection{Tractable problems}
\label{secscase}

If $\Xset$ is a polyhedron in $\Rset_+^n$, then~(\ref{lpf}) is an uncertain linear programming problem. In this case~(\ref{bnsmoduc1}), for a fixed $\lambda\in [0,1]$, is a system of linear constraints over $\Rset_+^n$, whose feasibility can be tested in polynomial time (see, e.g.,~\cite{SCH86}). In consequence, \textsc{Soft-Nec}~$\widetilde{\mathcal{P}}$ can be then solved in polynomial time with a fixed accuracy $\epsilon>0$. 

If the integrality assumptions on some variables are imposed or $\Xset\subseteq \{0,1\}^n$, then checking the feasibility of~(\ref{bnfsp}), for a fixed $\lambda\in [0,1]$, is NP-hard in general (see, e.g.,~\cite{GJ79}). We now describe a special case of such a problem, which can be solved efficiently. Consider the following combinatorial optimization problem with uncertain costs:
\begin{equation}
 \begin{array}{lll}
\min & x_0\\
 \text{s.t.}  & \tilde{\pmb{c}}^T\pmb{x}-x_0\leq 0 \\
 		& \pmb{x}\in \Xset \subseteq \{0,1\}^n
  \end{array}
 \label{lpcomb}
\end{equation}
Using (\ref{bnsmoduc1}), we can express~(\ref{lpcomb})  as follows:
\begin{equation}
\label{bnsmoducomb}
 \begin{array}{llll}
                         \max  &(1-\lambda) \\
                    \text{s.t.}      &  \multicolumn{2}{l}{\displaystyle \hat{\pmb{c}}^T \pmb{x}+\Gamma_0w_0+\sum_{j\in [n]}q_j\leq \hat{c}+\zeta(1-\lambda)+ \gamma_0(1-\lambda)} \\
                                      &w_0+q_{j}\geq \beta_j(\lambda)x_j  &   j\in [n] \\
                                      & w_0\geq 0 \\
                                      &0\leq \lambda\leq 1\\
                                       &\pmb{x}\in \Xset\subseteq \{0,1\}^n
 \end{array}
 \end{equation}
 where $\hat{c}=\min_{\pmb{x}\in \Xset} \hat{\pmb{c}}^T\pmb{x}$. Using similar relation as the one between~(\ref{robbert0}) and~(\ref{e0}), we can equivalently express~(\ref{bnsmoducomb}) as 
\begin{equation}
\label{e05}
	\begin{array}{llll}
		\max  & 1-\lambda \\
		\text{s.t.}  & \displaystyle \max_{\pmb{c}\in  \mathcal{S}_0 \cap  \mathcal{U}_0^\lambda} \pmb{c}^T\pmb{x}\leq \hat{c}+\zeta(1-\lambda)+ \gamma_0(1-\lambda)\\
		  &0\leq \lambda\leq 1\\
				 & \pmb{x}\in \Xset\subseteq \{0,1\}^n
	\end{array}
\end{equation}
where $\mathcal{S}_0$ and $\mathcal{U}_0^\lambda$ were defined in Section~\ref{secfobj}.
If $(\pmb{x}^*,\lambda^*)$ is an optimal solution to~(\ref{e05}), then $\pmb{x}^*$ is a best necessarily soft  feasible solution
with  $\mathrm{N}((\pmb{x}^*,x_0) \text{  is } \Gamma_0-\widetilde{\textsc{Feas}}) \wedge
 (x_0 \widetilde{\leq} \widetilde{C}))=1-\lambda^*$. 
A method of solving~(\ref{e05}) is based on a binary search in the interval of possible values of~$\lambda\in [0,1]$.
In order to test the feasibility of~(\ref{e05}) of a fixed $\lambda$, we can first solve the problem
\begin{equation}
\label{e06}
	\min_{\pmb{x}\in \Xset} \max_{\pmb{c}\in  \mathcal{S}_0 \cap  \mathcal{U}_0^\lambda} \pmb{c}^T\pmb{x}
\end{equation}
and check then if the optimal objective value of~(\ref{e06}) is not greater than $\hat{c}+\zeta(1-\lambda)+ \gamma_0(1-\lambda)$.
To solve~(\ref{e06}) we can use the algorithm proposed in~\cite[Theorem~1]{LK14}. 
It consists of solving $\lceil \frac{n-\Gamma}{2}\rceil+1$ deterministic counterparts of problem~(\ref{lpcomb}) in 
$\left(\lceil \frac{n-\Gamma}{2}\rceil+1\right)T(n)$ time, where  $T(n)$ is the time required to solve one deterministic problem.
The algorithm for solving~(\ref{e05})
is an adaptation of Algorithm~\ref{algbnfs}
 (it is enough to solve deterministic problem under the nominal costs $\hat{\pmb{c}}$ in Step~\ref{algbnfs5a}
and 
 apply the algorithm proposed in~\cite[Theorem~1]{LK14} in Step~\ref{algbnfs5}).
Its overall running time is now
 $O(\left(\lceil \frac{n-\Gamma}{2}\rceil+1\right) T(n)\lceil\log \epsilon^{-1} \rceil)$,  where $\epsilon>0$ is a  given accuracy
 and $\Gamma\leq n$. Therefore,
the algorithm is polynomial under the assumption that solving 
 the deterministic counterpart of problem~(\ref{lpcomb}) can be done in polynomial time. This is true for 
 such problems as:  shortest path, minimum spanning tree, minimum assignment,~etc.
(see, e.g.,~\cite{CO09,PS98}).

\section{Computational experiments}
\label{secexper}
In this section we show the results of some computational tests. Our goal is to compare the soft robust approach in the possibilistic setting, proposed in Section~\ref{secsoftrob}, to the concept of light robustness presented in~\cite{FM09, S14}.
We examine uncertain linear programming problem of the following form:
\begin{equation}
 \begin{array}{lll}
\min &\pmb{c}^T\pmb{x}\\
 \text{s.t.} &\tilde{\pmb{a}}_i^T\pmb{x}\leq b_i & i\in [m]\\
 		& \pmb{x}\in [0,1]^n
  \end{array}
 \label{lpf01}
\end{equation}
We assume that the objective function is deterministic (only the constraints are uncertain). An instance $I$ of the problem~(\ref{lpf01}) is generated as follows:
\begin{enumerate}
	\item the number of variables $n=100$ and the number of constraints $m=5$;\label{p1}
	\item each cost $c_j$, $j\in [n]$, is a random integer, uniformly distributed in the interval $[-100,-1]$;
	\item the nominal value of the constraint coefficient $\hat{a}_{ij}$ is a random integer, uniformly distributed in the interval $[1,100]$ and the bound $\overline{a}_{ij}$ is set to $\sigma\cdot \hat{a}_{ij}$, where $\sigma$ is a random number uniformly distributed in the interval $[0,1]$;  
	\item we fix $b_i=0.3\sum_{j\in [n]}\hat{a}_{ij}$ for each $i\in [m]$. \label{p4}
\end{enumerate}
 We set the protection levels $\Gamma_i=30$ for each $i\in [m]$. In the light robustness concept (see model~(\ref{lplr})) we use the $||\pmb{\gamma}||_{\infty}=\max\{\gamma_1,\dots,\gamma_m\}$ norm. In the soft robust approach (see model~(\ref{bnsmod})) we assume the $10\%$ tolerance for the constraint violation, i.e. $\overline{b}_i=0.1b_i$ for each $i\in [m]$.  For the membership functions of all fuzzy sets we fix $z=1$, so their membership functions are piecewise linear. In particular, the uncertain coefficients  $\tilde{a}_{ij}$ are triangular fuzzy intervals. Let $\hat{c}$ be the optimal objective value of the deterministic counterpart of~(\ref{lpf01}) under the nominal scenario $\widehat{\pmb{A}}$. We will choose $\rho_0=p\cdot|\hat{c}|$ for $p\in \{0, 0.2\%, 0.4\%,\dots,10\%\}$.
 
Let $\pmb{x}\in \Xset$ be a solution to~(\ref{lpf01}), obtained by solving the model~(\ref{bnsmod}). We will compute the distance of $\pmb{x}$ to the optimum under the nominal scenario as follows:

 $$d(\pmb{x})=\left|\frac{\pmb{c}^T\pmb{x}-\hat{c}}{\hat{c}}\right|.$$
 The value of  $d(\pmb{x})$ is the price of robustness of $\pmb{x}$.  In order to evaluate the \emph{a posteriori} quality of~$\pmb{x}$ we use the following  Monte Carlo simulation. For each coefficient $\tilde{a}_{ij}$, independently, we generate its value (realization) as follows. First we choose uniformly at random $\lambda \in [0,1]$ and then uniformly at random the realization $a_{ij}\in \tilde{a}_{ij}^\lambda=[\hat{a}_{ij}-\overline{a}_{ij}(1-\lambda),\hat{a}_{ij}+\overline{a}_{ij}(1-\lambda)]$. Observe that realizations closer to $\hat{a}_{ij}$ are more probable.
  This gives us a scenario $\pmb{A}=(a_{ij})\in \Rset_{+}^{m\times n}$, which provides a deterministic counterpart of~(\ref{lpf01}). For this deterministic problem we compute the magnitude of the constraint violation of $\pmb{x}$, i.e. the value ${\rm viol}(\pmb{x},\pmb{A})=\max_{i\in [m]} [(\pmb{a}_i^T\pmb{x}-b_i)/b_i]^+$,
  where $[y]^+=\max\{0,y\}$.
   After generating a set $\mathbb{A}$ of $1000$ random scenarios, we computed the fraction of the scenarios under which $\pmb{x}$ is infeasible, i.e. 
   $$\#{\rm infeas}(\pmb{x})=\frac{|\{\pmb{A}\in \mathbb{A}: {\rm viol}(\pmb{x}, \pmb{A})>0\}|}{1000}$$ and the average magnitude of the constraint violation $${\rm aviol}(\pmb{x})=\frac{1}{1000}\sum_{\pmb{A}\in \mathbb{A}} {\rm viol}(\pmb{x}, \pmb{A}).$$
The quantities $d(\pmb{x})$, $\#{\rm infeas}(\pmb{x})$ and ${\rm aviol}(\pmb{x})$ can be seen as \emph{a posteriori} evaluation of the quality of $\pmb{x}$.

  The experiments were performed as follows. For each  $p\in \{0, 0.2\%, 0.4\%,\dots,10\%\}$ we generated 100 instances $I_1,\dots, I_{100}$ as shown in points~\ref{p1}-\ref{p4}. For each instance $I_i$ we fixed $\rho_0=p\cdot \hat{c}_i$ and computed an optimal light robust solution~$\pmb{x}^L_i$, by solving~(\ref{bsmodl}), and a best necessarily soft feasible solution~$\pmb{x}^S_i$, by solving~(\ref{bnsmod}). 
  For solving the models~(\ref{bsmodl}) and~(\ref{bnsmod})  we used  \texttt{IBM ILOG CPLEX 12.9} 
  optimizer~\cite{CPLEX}
and the modeling package \texttt{JuMP}~\cite{LD15}
embedded in the programming language \texttt{Julia}.
 We computed the average qualities of the solutions. Namely, the average qualities of optimal light robust solutions are
  $$d^L(p)=\frac{1}{100}\sum_{i\in [100]} d(\pmb{x}^L_i),$$
  $$\#{\rm infeas}^L(p)=\frac{1}{100}\sum_{i\in [100]} \#{\rm infeas}(\pmb{x}^L_i),$$
  $${\rm aviol}^L(p)=\frac{1}{100}\sum_{i\in [100]} {\rm aviol}(\pmb{x}^L_i).$$
  The value of $\#{\rm infeas}^L(p)$ can be interpreted as the fraction of 100~000 deterministic counterparts for which an optimal light robust solution was infeasible (at least one constraint was violated) for a fixed~$p$. Accordingly, the value of ${\rm aviol}^L(p)$ is the average magnitude of the infeasibility. The quantities $d^S(p)$,  $\#{\rm infeas}^S(p)$ and ${\rm aviol}^S(p)$ for the set of best necessarily soft feasible solutions are computed in the same way. 
    \begin{figure}[ht]
	\centering
	\includegraphics[height=5.5cm]{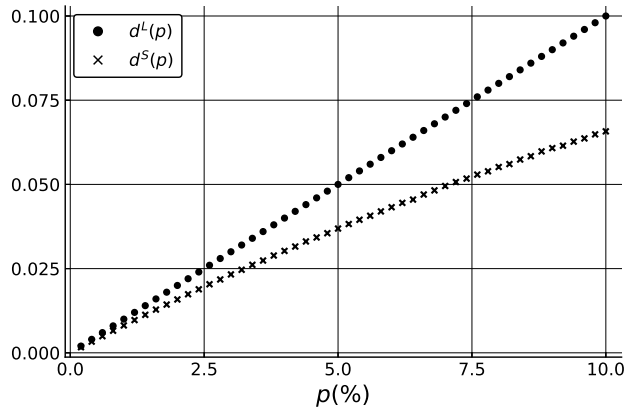}
	\caption{Average prices of robustness for various $p=\rho_0/|\hat{c}|$}\label{fig7}
\end{figure}
 \begin{figure}[ht]
	\centering
	\includegraphics[height=5.5cm]{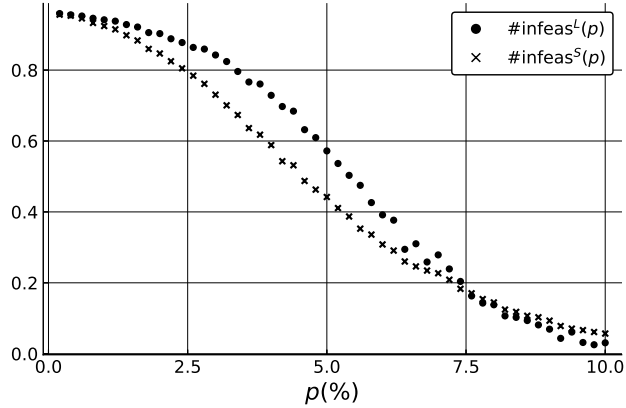}
	\caption{Fractions of infeasible solutions for various $p=\rho_0/|\hat{c}|$. }\label{fig4}
\end{figure}

Figure~\ref{fig7} shows the average prices of robustness of the computed solutions for various ratios $p=\rho_0/|\hat{c}|$. One can observe that $\pmb{x}^S$ have smaller prices of robustness than $\pmb{x}^L$. Furthermore, the difference between the prices becomes greater for larger $p$. This observation can be explained as follows.  In model~(\ref{bsmodl}) we use the constraint $\pmb{c}^T\pmb{x}\leq \hat{c}+\rho_0$, which is tight at the optimum. So, the figure of $d^L(p)$ is linear. In contrast, in the model~(\ref{bnsmod}) we use the flexible constraint, which yields $\pmb{c}^T\pmb{x}\leq \hat{c}+\zeta(1-\lambda)=\hat{c}+\lambda\rho_0$. Because, $\lambda\in [0,1]$, the cost of the  solutions $\pmb{x}^S$ can be closer to $\hat{c}$.

\begin{figure}[ht]
	\centering
	\includegraphics[height=5.5cm]{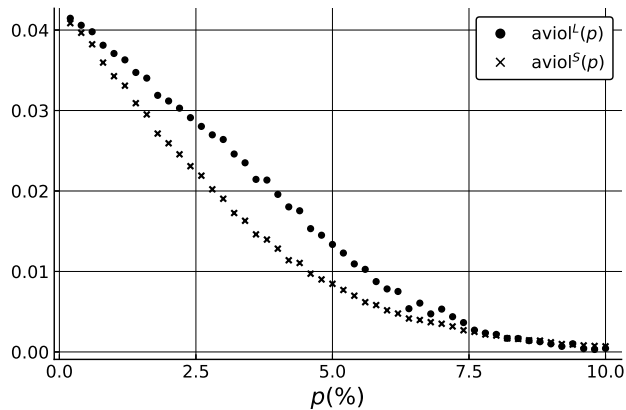}
	\caption{Average magnitudes of infeasibility for various $p=\rho_0/|\hat{c}|$.}\label{fig5}
\end{figure}

Figures~\ref{fig4} and~\ref{fig5} show the fractions of infeasible solutions and the average magnitude of constraints violations for both tested approaches. If $p=0$, then both $\pmb{x}^S$ and $\pmb{x}^L$ must be optimal under $\hat{\pmb{c}}$ (their prices of robustness equal~0). In this case they robustness is very weak, i.e. almost all deterministic counterparts are infeasible. 
 Increasing $p$ (equivalently, the tolerance $\rho_0$), we can improve the robustness of both $\pmb{x}^S$ and $\pmb{x}^L$. For $p\geq 10\%$ almost all deterministic counterparts are feasible. However, the average price of robustness of $\pmb{x}^L$ is 0.1 whereas the  average price of robustness of $\pmb{x}^L$ is about 0.06. For $p\in (0,7.5\%)$ the solutions $\pmb{x}^S$ are more robust than $\pmb{x}^L$, have smaller average magnitude of the constraints violation and also have 
 a
 smaller price of robustness. We can thus conclude that taking the possibilistic information into account can improve the quality of the obtained solutions.

  \section{Conclusions}
  
  In this paper we have proposed a new concept of choosing a solution in uncertain
  optimization problems,  in which unknown
parameters are modeled by fuzzy intervals whose membership functions are regarded as
 possibility distributions for their values.
   In the traditional robust approach the values of uncertain parameters are only known to belong to a given uncertainty set $\mathcal{U}$.  We then seek a solution which behaves reasonably under the worst parameter realizations in $\mathcal{U}$. This traditional robust approach has some well-known drawbacks. It does not take any additional information connected with $\mathcal{U}$ into account. Furthermore, it is often considered to be too pessimistic (conservative) as the probability of occurrence of bad scenarios may be small. Our approach overcome these drawbacks. By specifying the possibility distribution in $\mathcal{U}$, as an upper bound on the unknown probability distribution, we provide additional information which can be utilized to improve the quality of  computed solutions. Furthermore, following the idea of light robustness, we allow decision makers to control the price of robustness of the solutions. It is important that the proposed model can be solved in polynomial time if the underlying deterministic counterpart is polynomially solvable. In particular, this is true for uncertain linear programming problems and some
   uncertain combinatorial optimization problems
   (shortest path, minimum spanning tree, minimum assignment, etc.)
   
\subsubsection*{Acknowledgements}
This work was  supported by
 the National Science Centre, Poland, grant 2017/25/B/ST6/00486.


\appendix

\section{Appendix}
\label{dod}  

In this appendix we show the transformation from~(\ref{robbert}) to (\ref{e0}) originally obtained in~\cite{BS04}.
Fix $\pmb{x}\in \Xset$ and consider the $i$th constraint~(\ref{robbert}). 
Since the first term in~(\ref{robbert}) is fixed,
we focus on the following
optimization problem over the set of constraint coefficient realizations $\mathcal{S}_i\cap \mathcal{U}_i$,
namely
\begin{equation}
\label{maxci}
 \max_{\{N_i\subseteq [n]: |N_i|\leq\Gamma_i\}}\sum_{j\in N_i} \overline{a}_{ij} x_j.
 \end{equation}
Problem~(\ref{maxci}) can be formulated by the following linear programming problem:
\begin{equation}
  \label{maxcilp}
	\begin{array}{llll}
			\max & \displaystyle  \sum_{j\in [n]} (\overline{a}_{ij} x_j) \delta_{ij} &&\\
			\text{s.t.} &  \displaystyle  \sum_{j\in [n]} \delta_{ij}\leq \Gamma_i&& \braket{w_i}\\
 					& \delta_{ij}\leq 1&j\in[n]& \braket{p_{ij}}\\
					&\delta_{ij}\geq 0&j\in[n]&
		\end{array}
\end{equation}
Indeed,  the constraint matrix of problem~(\ref{maxcilp}) is \emph{unimodular}  and each vertex solution~$\pmb{\delta}$ is such that
$\pmb{\delta}\in \{0,1\}^n$ (see, e.g.,~\cite{PS98}). 
Hence these decision variables express  the selection of subset~$N_i\subseteq [n]$.
Since $\overline{a}_{ij} x_j\geq 0$, an optimal solution consists of~$\Gamma_i$ variables at~$1$.
The dual of problem~(\ref{maxcilp}) is as follows (the dual variables corresponding to the constraints in~(\ref{maxcilp})
are in the brackets):
\begin{equation}
 \label{dmaxcilp}
	\begin{array}{lllll}
		\min &\displaystyle\Gamma_i w_i+\sum_{j\in [n]} p_{ij}\\
		\text{s.t.} &w_i+p_{ij}\geq \overline{a}_{ij}x_j & j\in [n]\\
		&w_i\geq 0, p_{ij}\geq 0 & j\in [n]
	\end{array}
\end{equation}
Clearly problem~(\ref{maxcilp}) (problem~(\ref{maxci}))  is feasible and bounded for all integer $\Gamma_i$ in $[0,n]$.
By strong  duality (see, e.g.,~\cite{PS98}), problem~(\ref{dmaxcilp}) is  feasible and bounded  as well. At optimality the values of their
objective functions are equal.
Replacing (\ref{maxci}) by (\ref{dmaxcilp}) in~(\ref{robbert}), we have
\[
	\begin{array}{lllll}
		\displaystyle\hat{\pmb{a}}_i^T\pmb{x}+\Gamma_i w_i+\sum_{j\in [n]} p_{ij} \leq b_i\\
		w_i+p_{ij}\geq \overline{a}_{ij}x_j & j\in [n]\\
		w_i\geq 0, p_{ij}\geq 0 & j\in [n]
	\end{array}
\]
Hence (\ref{robbert}) is equivalent to~(\ref{e0}).

\end{document}